\pdfoutput=1
\documentclass[11pt]{article}

\usepackage[final]{acl}

\usepackage{times}
\usepackage{latexsym}
\usepackage{amsmath}

\usepackage[T1]{fontenc}
\usepackage[utf8]{inputenc}

\usepackage{microtype}

\usepackage{times}  % DO NOT CHANGE THIS
\usepackage{graphicx} % DO NOT CHANGE THIS
\usepackage{natbib}  % DO NOT CHANGE THIS AND DO NOT ADD ANY OPTIONS TO IT
\usepackage{diagbox}
\usepackage{caption} % DO NOT CHANGE THIS AND DO NOT ADD ANY OPTIONS TO IT
\usepackage[utf8]{inputenc}
\usepackage[T1]{fontenc}
\usepackage[english]{babel}
\usepackage{amsfonts}
\usepackage{nicefrac}
\usepackage{microtype}
\usepackage{amsmath}
\usepackage{amssymb}
\usepackage{mathtools}
\usepackage{subcaption}
\usepackage{multirow}
\usepackage{booktabs}
\usepackage{ragged2e}
\usepackage{tikz}
\usepackage{stackengine}
\usepackage{etoolbox}
\usepackage{xspace}
\usepackage{xpatch}
\usepackage{cuted}
\usepackage{enumerate}
\usepackage{xstring}
\usepackage{setspace}
\usepackage{tabularx}
\usepackage{makecell}
\usepackage{changepage}
\usepackage{enumitem}
\usepackage{cuted}
\usepackage{titlesec}
\usepackage{cancel}
\usepackage{eqparbox}
\usepackage{bibentry}
\usepackage{algpseudocode}
\usepackage[hang,flushmargin]{footmisc}
\usepackage[capitalise,noabbrev,nameinlink]{cleveref}
\usepackage[useregional=numeric]{datetime2}
\usepackage[linesnumbered,ruled,noend,noline]{algorithm2e}
\usepackage{adjustbox}

\SetKwInput{KwModel}{Model}
\SetKwInput{KwParameter}{Parameter}
\SetKw{KwBy}{by}
\SetKw{KwIn}{in}
\SetKwInput{KwMetric}{Metric}

\usepackage[switch,mathlines]{lineno}
\usepackage{lipsum}% http://ctan.org/pkg/lipsum
\usepackage{etoolbox}
\usepackage{longtable}
\newtheorem{definition}{Definition}
\newtheorem{theorem}{Theorem}
\newtheorem{proposition}{Proposition}

\makeatletter
\def\@opargbegintheorem#1#2#3{\trivlist
   \item[]{\textbf{#1\ #2} \ (#3)\textbf{.}\ } \itshape}
\makeatother

\graphicspath{{figures/}}

\usepackage{xcolor}

\title{Certified Error Control of Candidate Set Pruning for \\Two-Stage Relevance Ranking}

\author{Minghan Li\thanks{\quad Equal contribution},
        Xinyu Zhang\footnotemark[1],
        Ji Xin,
        Hongyang Zhang,
        Jimmy Lin\\[1ex]
        David R. Cheriton School of Computer Science, University of Waterloo\\[1ex]
        \texttt{\{m692li,x978zhang,ji.xin,hongyang.zhang,jimmylin\}@uwaterloo.ca}}

\begin{document}
\maketitle
\begin{abstract}
In information retrieval (IR), candidate set pruning has been commonly used to speed up two-stage relevance ranking. 
However, such an approach lacks accurate error control and often trades accuracy off against computational efficiency in an empirical fashion, lacking theoretical guarantees.
In this paper, we propose the concept of \textit{certified error control} of candidate set pruning for relevance ranking, which means that the test error after pruning is guaranteed to be controlled under a user-specified threshold with high probability.  
Both in-domain and out-of-domain experiments show that our method successfully prunes the first-stage retrieved candidate sets to improve the second-stage reranking speed while satisfying the pre-specified accuracy constraints in both settings.
For example, on MS MARCO Passage v1, our method yields an average candidate set size of 27 out of 1,000 which increases the reranking speed by about 37 times, while the MRR@10 is greater than a pre-specified value of 0.38 with about 90\% empirical coverage and the empirical baselines fail to provide such guarantee.
Code and data are available at: \url{https://github.com/alexlimh/CEC-Ranking}.
\end{abstract}

 \section{Introduction}
The two-stage relevance ranking architecture has been an indispensable component for knowledge-intensive natural language processing tasks such as information retrieval (IR)~\citep{Manning2005IntroductionTI} and open-domain question answering (OpenQA)~\citep{chen-etal-2017-reading}.
Such a system usually consists of a high-recall retriever that retrieves a set of documents from a massive corpus and a high-precision reranker that improves the ranking of the retrieved candidate sets.
The first-stage retrieval, often implemented by approximate nearest neighbour search~\citep{johnson2017faiss} or inverted index search~\citep{lin2021pyserini}, is quite efficient while the second stage reranking usually has high latency due to the trend of using over-parameterized pre-trained language models and large candidate set size.
Previous work in early exiting proposed to predict the ranking score using only a partial model~\citep{xin-etal-2020-early,Lucchese2020QuerylevelEE,Busolin2021LearningEE} or prune the candidate set before reranking~\citep{wang2011cascade,Fisch2021EfficientCP} to trade accuracy off against speed.
However, such methods lack accurate error control which fails to satisfy the exact accuracy constraints specified by users.

\begin{figure}
    \centering
    \includegraphics[width=0.45\textwidth]{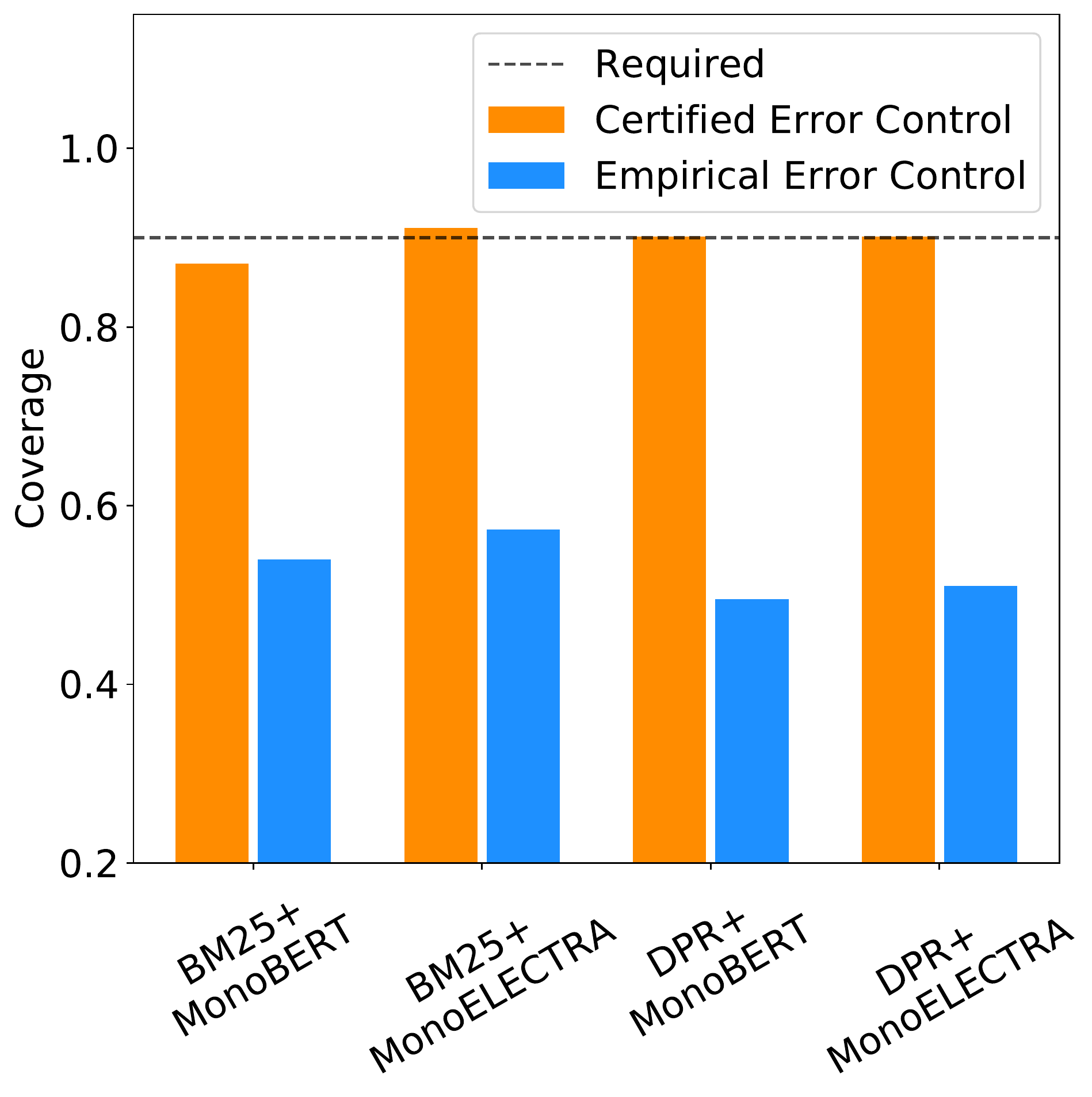}
    \caption{Comparison between certified and empirical error control methods on MS MARCO Passage v1. Coverage: Percentage of 100 independent runs that satisfies a pre-specified MRR@10$\geq $0.35  (the dotted line).}
    \label{fig:demo}
\end{figure}

\begin{figure*}[t!]
    \centering
    \includegraphics[width=0.83\textwidth]{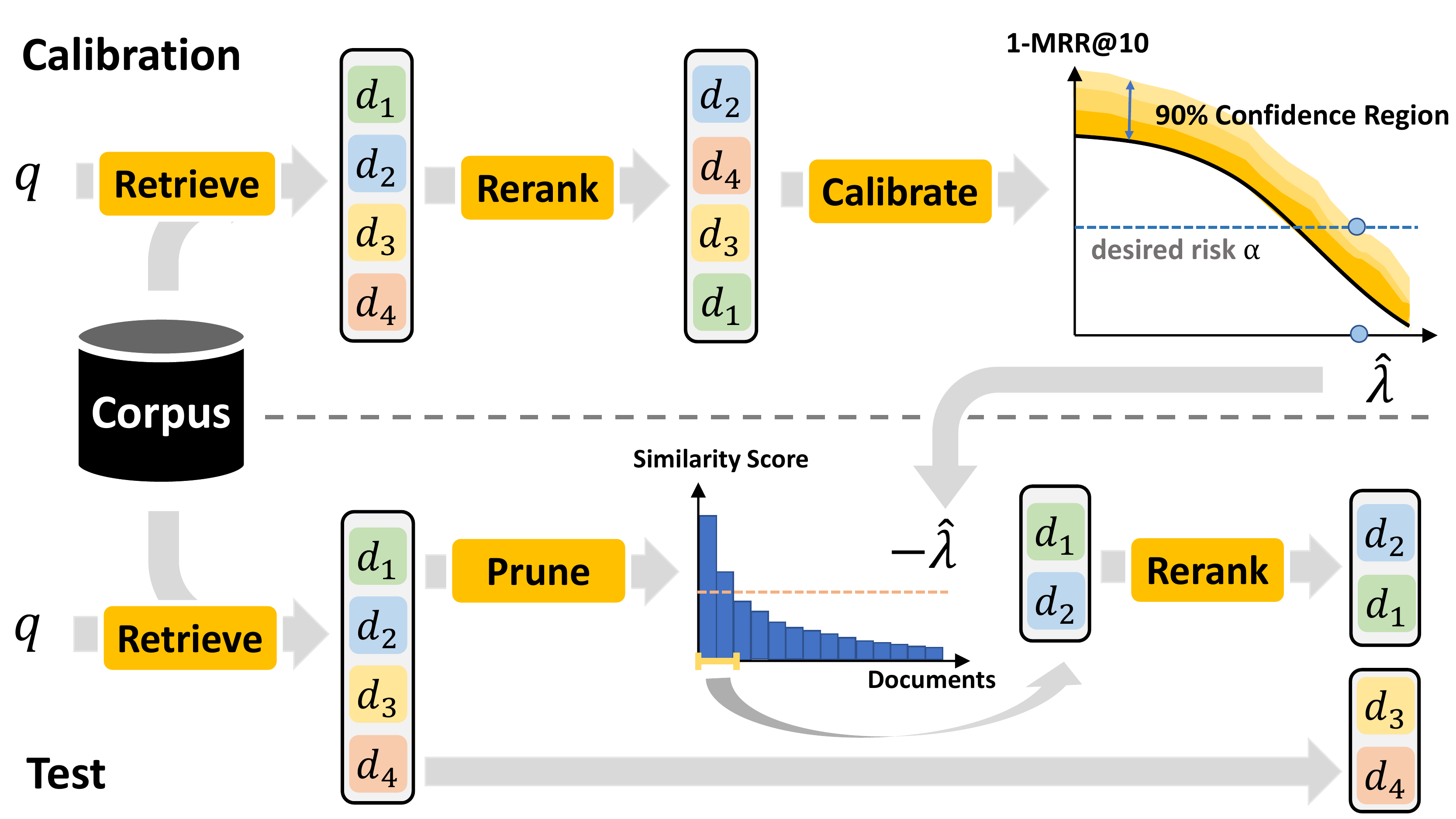}
    \caption{Calibration and test procedures of candidate set pruning with certified error control. During calibration, we use the confidence region and desired risk $\alpha$ to find the threshold $-\hat{\lambda}$. During testing, we use the threshold to prune the retrieved documents, which guarantees to control the expected loss under $\alpha$ with probability $1-\delta$.}
    \label{fig:rcps}
\end{figure*}

In this paper, we focus on the \textit{candidate set pruning} methods of early exiting for two-stage relevance ranking, and we show that a simple score-thresholding method could yield predictions with \textit{certified} error control using the prediction sets theory~\citep{Wilks1941DeterminationOS,Wilks1942StatisticalPW,Wald1943AnEO, Bates2021DistributionFreeRP}.
Instead of predicting a single target, prediction sets will yield sets that contain the desired fraction of the population with high probability.
Moreover, we allow users to specify the error tolerance of their custom metrics and our method will return the pruned set of candidates that satisfies the constraints with a finite sample guarantee.
Our method makes no assumption about the data distributions and models, except that the calibration data are exchangeable with the test data.
Fig.~\ref{fig:rcps} illustrates the calibration and test procedure for certified error control of candidate set pruning.

\medskip
\noindent\textbf{Challenges.}
However, directly applying the prediction set methods to relevance ranking could be problematic.
Unlike classification where we mainly care about whether the true label is in the predicted set (i.e., recall), in relevance ranking, the evaluation metric relates to the rank of the true positive in the set.
Therefore, it would be unreasonable to set a high standard for ranking quality if the ranking system has poor performance, as the average rank of the true positive documents might not reach the requirement no matter how large the set is.

\medskip
\noindent\textbf{Contributions.}
To this end, we propose to correct the risk threshold or confidence level if the constraints are impossible to satisfy.
Specifically, we will adjust the risk threshold to the minimum of the risk's upper bound or decrease the confidence level to shrink the upper bound.
Finally, it is up to users to decide whether to abstain the results or accept the corrected results.
Empirically, we evaluate our method on IR and ODQA benchmarks, including MS MARCO Passage v1~\citep{nguyen2016ms} and Quora~\citep{thakur2021beir}.
We also test different combinations of retrievers and rerankers for the two-stage relevance ranking system under both in-domain and out-of-domain settings.
Fig.~\ref{fig:demo} shows the results of certified and empirical error control methods using different ranking systems on MS MARCO Passage v1, and we could see that the empirical fails to provide the required coverage while our method succeeds (see Tbl.~\ref{tbl:marco_coverage} for more details).
For example, if we pre-specify MRR@10 $\geq$ 0.38, we could yield an average candidate set size 27 out of 1,000, increasing the reranking speed by 37$\times$ while satisfying the constraint with 90\% empirical coverage.
We further confirm that the risk-confidence correction of our method is able to consistently correct the risk/confidence when the pre-specified conditions are impossible to achieve.

To sum up, our contributions are three-fold:
\begin{itemize}
    \item We propose the first certified error control method of candidate set pruning for relevance ranking, providing a guarantee to satisfy user-specified constraints with high probability.
    \item We propose a risk-confidence correction method to adjust the constraints otherwise impossible to satisfy for ranking tasks.
    \item Our method has consistent results under both in-domain and out-of-domain settings. With at least 90\% coverage, our method returns candidates less than 50 with $1\sim2\%$ accuracy drop for most ranking systems. 
\end{itemize}

\section{Related Work}\label{sec:related}

\noindent\textbf{Early exiting for relevance ranking.}
Early exiting~\cite{deebert,fastbert,xin-etal-2021-berxit} is a popular latency-accuracy tradeoff method in document ranking.
\citet{xin-etal-2020-early} and \citet{soldaini-moschitti-2020-cascade} proposed to output the question-document similarity score at earlier layers of a pre-trained language models, while \citet{Lucchese2020QuerylevelEE} and \citet{Busolin2021LearningEE} proposed to use a set of decision trees in the ensemble for prediction.
\citet{Cambazoglu2010EarlyEO} proposed optimization strategies that allow short-circuiting score computations in additive learning systems.
\citet{wang2011cascade} presented a boosting algorithm for learning such cascades to optimize the tradeoff between effectiveness
and efficiency.
Despite their popularity, the above early exiting methods mainly use fixed rules for efficiency-accuracy tradeoffs without performance guarantees.

\smallskip
\noindent\textbf{Prediction sets and cascade systems.}
Prediction sets are essentially tolerance regions ~\cite{Wilks1941DeterminationOS,Wilks1942StatisticalPW,Wald1943AnEO}, which are sets that contain the desired fraction of the collection with high probability.
Recently, tolerance regions have been applied to yield prediction sets for deep learning models~\citep{Park2020PACCS,Park2021PACCP}. 
~\citet{Bates2021DistributionFreeRP}. 
In addition, conformal prediction~\citep{Vovk1999MachineLearningAO,10.5555/1062391} has been recognized as an attractive way of producing predictive sets with finite-sample guarantees.
In retrieval, structured prediction cascades~\citep{Weiss2010StructuredPC} optimize their cascades for overall pruning efficiency, and \citet{Fisch2021EfficientCP} proposed a cascade system to prune the unnecessarily large conformal prediction sets for OpenQA.
However, conformal prediction is only suitable for metrics like recall but not for ranking-oriented tasks.

\section{Background}

\subsection{Notation}
\label{sec:background:setting}

In the rest of the paper, we will use upper-case letters (e.g., $Q$, $D$,...) to denote the random variables, script letters (e.g., $\mathcal{Q}, \mathcal{D}$,...) to denote the event space, and lower-case letters (e.g., $q$, $d$,...) to denote the implement of a random variable in its event space.
Specially, we use $\mathcal{X'}$ to denote the space of all possible subsets in $\mathcal{X}$ where $\mathcal{X'} = 2^{\mathcal{X}}$.

\subsection{Two-Stage Relevance Ranking}
Given a question $q$, the relevance ranking task is to return a sorted list of documents from a large text corpus to maximize a metric of interest.
Particularly, in a two-stage ranking system, a first-stage retriever generates a set of candidate documents $\{d_1, d_2, . . ., d_{k}\}$ for the second-stage reranker to re-order the candidate lists.

\subsubsection{Retrieval}

\noindent\textbf{Dense Retrievers} encode the question and documents separately and project them into a low-dimensional space (e.g., 768 dim, which is ``lower'' than the size of the corpus vocabulary). Representative methods include DPR~\citep{karpukhin-etal-2020-dense}, ANCE~\citep{xiong2020ance}, ColBERT~\citep{Khattab2020ColBERTEA}, and MeBERT~\citep{Luan2021SparseDA}.

\smallskip
\noindent\textbf{Lexical/Sparse Retrievers} use the corpus vocabulary as the basis for vector representation. 
Static methods include BIM~\citep{Manning2005IntroductionTI},  tf-idf~\citep{Ramos2003UsingTT}, and BM25~\citep{robertson2009bm25}. 
Contextualized methods include SPLADE~\citep{Formal2021SPLADESL}, DeepCT~\citep{Dai2019ContextAwareST}, DeepImpact~\citep{Mallia2021LearningPI}, and  COIL~\citep{gao-etal-2021-coil,Lin2021AFB}.

\medskip
\noindent Despite their difference, all the above methods could be viewed as a logical scoring model~\citep{Lin2021APC}.
Let $\eta_Q: \mathcal{Q}\rightarrow \mathbb{R}^n$ be a random function that maps a question to an $n$ dimensional vector representation, and let $\eta_D: \mathcal{D}\rightarrow \mathbb{R}^n$ be a random function that maps a document to an $n$ dimensional vector.
The similarity score $s_v$ between a question $q$ and a document $d$ can be defined as:
\begin{align}\label{eq:sim_rtv}
    s_{v}(q, d) \doteq  \phi(\eta_q (q), \eta_d (d)),
\end{align}
where $\phi$ is a metric that measures the similarity between encoded vectors of $\eta (q)$ and $\eta (d)$, such as dot product or cosine similarity.

\subsubsection{Reranking}
The reranker module is responsible for improving the ranking quality of the
candidate documents returned from the first-stage retrievers.
We focus on recent neural rerankers based on pre-trained language models such as MonoBERT~\citep{Nogueira2019MultiStageDR} and MonoELECTRA.
The reranker could also be seen as a logical scoring model.
However, instead of using a bi-encoder structure, a cross-encoder structure is often applied where the question-document pairs are encoded and fed into a single model together for more fine-grained token-level interactions:
\begin{align}\label{eq:sim_rrk}
    s_{r}(q, d) \doteq  \zeta (\text{concat}(q, d)),
\end{align}
where $\zeta$ is the reranker that takes the question-document pair as input and outputs the similarity score $s_r$.
One way to implement the ``concat'' function is using special tokens as indicators, such as $[\text{CLS}]\ q \ [\text{SEP}]\ d \ [\text{SEP}]$.

\section{Methods}\label{sec:methods}
The goal is to speed up the second-stage reranking by pruning the first-stage candidate sets, while the final error (e.g., 1-$\text{MRR@}10$) can be controlled under a user-specified level with high probability inspired by~\citet{Bates2021DistributionFreeRP}.

\subsection{Settings}\label{sec:rcps_ranking}
Formally, given a question $Q \in \mathcal{Q}$, a set of documents $D'\in\mathcal{D'}$ retrieved by the first-stage retriever $\phi$, and a relevance-judged set of gold documents $D'_\omega\in\mathcal{D'_\omega}$, we consider a pruning function $\mathcal{T} : \mathcal{D'} \rightarrow \mathcal{P'}$, where $\mathcal{P'}$ denotes the space of subsets of $D'$. 
We then use a loss function on the pruned document sets that also depends on the reranker $\zeta$, i.e., $L(\cdot, \cdot;\zeta):\mathcal{D'_\omega}\times\mathcal{P'} \rightarrow \mathbb{R}$, to encode a metric of the user's interest, and seek a pruning function $\mathcal{T}$ that controls the risk (i.e., error) $R(\mathcal{T};\zeta) = \mathbb{E}[L(D'_\omega, \mathcal{T}(D');\zeta)]$.

\begin{definition}[Certified Error Control of Candidate Set Pruning]\label{def:rcps}
Let $\mathcal{T}:\mathcal{D'} \rightarrow \mathcal{P'}$ be a random function. 
We say that $\mathcal{T}$ is a
pruning function for reranker $\zeta$ with certified error control if, with probability at least $1 - \delta$, we have $R(\mathcal{T};\zeta) \leq \alpha$.
\end{definition}
The risk level $\alpha> 0$ is pre-specified by users, and the same goes for $\delta\in(0,1)$ where $0.1$ is often chosen as a rule of thumb.

\subsubsection{Pruning Function and Risk Function}
We use a calibration set to certify the error control of the pruning function and apply the certified pruner during testing.
Let $\{Q_i
, D'_i, {D'_\omega}_i\}_{i=1}^m$ be an i.i.d. sampling set of random variables representing a calibration set of queries, candidate document sets, and gold document sets.
For the pruning function, we define a parameter $\lambda\in\Lambda$ as its index which is essentially a score threshold with the following property:
\begin{align}\label{eq:set_predictor}
    \lambda_1 < \lambda_2 \Rightarrow \mathcal{T}_{\lambda_1}(d') \subset \mathcal{T}_{\lambda_2}(d').
\end{align}
Let $L(D'_\omega, P'; \zeta): \mathcal{D'_\omega}\times\mathcal{P'} \rightarrow \mathbb{R}_{\geq 0}$ be a loss function on the pruned  subsets. 
In ranking, we could take, for instance, $L(D'_\omega, P'; \zeta) = 1 - \text{MRR@10}(D'_\omega, P'_{\zeta})$, where MRR@10 is a classical measurement of ranking quality and $P'_{\zeta}$ is the reranked version $P'$ by $\zeta$. 
In general, the loss function has the following nesting property:
\begin{align}\label{eq:monotonicity}
 P'_1 \subset P'_2 \Rightarrow L(D'_\omega, P'_1; \zeta) \geq L(D'_\omega, P'_2; \zeta).
\end{align}
That is, larger sets lead to smaller loss (i.e., monotonicity)~\citep{Bates2021DistributionFreeRP}.
We then define the risk of a pruning function $\mathcal{T}_{\lambda}$ to be
\begin{align}
    R(\mathcal{T_{\lambda}}; \zeta) = \mathbb{E}[L(D'_\omega, P'; \zeta)].\nonumber
\end{align}

\subsubsection{Confidence Region}
In practice, to find the parameter $\lambda$, we need to search across the collection of functions $\{\mathcal{T}_{\lambda}\}_{\lambda\in\Lambda}$ and estimate their risk on the calibration set.
However, the true risk is often unknown and the empirical risk function is often used as an approximation:
\begin{align}
    \widehat{R}(\mathcal{T}_\lambda;\zeta) = \frac{1}{m}\sum_{i=1}^m L(D'_{\omega i}, \mathcal{T}_\lambda(D'_i); \zeta).\nonumber
\end{align}
To compute the confidence region, we leverage the concentration inequalities and assume that we have access to a pointwise confidence region for the risk function for each $\lambda$:
\begin{align}\label{eq:ucb}
    \Pr(R(\mathcal{T_{\lambda}}; \zeta) \leq \widehat{R}_{\delta}^+(\mathcal{T_{\lambda}}; \zeta)) \geq 1-\delta,
\end{align}
where $\widehat{R}(\mathcal{T_{\lambda}})$ is the empirical risk that estimated from $\{Q_i
, D'_i, {D'_\omega}_i\}_{i=1}^m$ and $\widehat{R}_{\delta}^+(\mathcal{T_{\lambda}}; \zeta)$ is its upper bound given a specific ${\delta}$ value.
\citet{Bates2021DistributionFreeRP} presented a generic strategy to obtain such bounds
by inverting a concentration inequality as well as concrete bounds for various settings.
For this paper, we use the Waudby-Smith-Ramdas (WSR) bound~\citep{WaudbySmith2020VarianceadaptiveCS} which is adaptive to variance.
We provide the specific form of WSR bound in Appendix~\ref{sec:appendix:wsr}.

We choose the smallest $\lambda$ such that the entire confidence region to the right of $\lambda$ falls below the target risk level $\alpha$ following~\cite{Bates2021DistributionFreeRP}:
\begin{align}\label{eq:lammbda_hat}
\hat{\lambda} \doteq \text{inf}\left\{ \lambda \in \Lambda:\widehat{R}_{\delta}^+(\mathcal{T_{\lambda'}}; \zeta)<\alpha, \forall \lambda'\geq\lambda \right\}.
\end{align}

\noindent In this way, $\mathcal{T}_{\hat{\lambda}}$ is a pruning function with certified error control. 
Theorems and proofs are provided in the Appendix~\ref{sec:appendix:proofs}.
In the following sections, we will discuss the problems of truncated risk function in relevance ranking and how to modify the certification for impossible constraints.

\subsection{Truncated Risk Function for Ranking}\label{sec:risk_func}
In Section~\ref{sec:rcps_ranking}, we mentioned that the risk function is often related to the metrics that we care about.
For example, in ranking, metrics such as MRR@K are often used to assess the ranking quality, where K is the maximal set size that we choose.
For example, the MRR@10 score for a set of questions $\{q_i\}_{i=1}^m$, the positive document sets $\{d'_i\}_{i=1}^m$, and the retrieved document candidate sets $\{d'_{\omega i}\}_{i=1}^m$ are
\begin{align}
    \text{MRR@10} = \frac{1}{m}\sum_{i=1}^{m} \frac{1}{f(d'_{i}, d'_{\omega i})},
\end{align}
where
\begin{align}\label{eq:mrr_threshold}
    f(d'_{i}, d'_{\omega i}) = \left\{\begin{matrix}
+\infty, & \text{if } r_i > 10; \\ 
r_i, &  \text{otherwise},
\end{matrix}\right.
\end{align}
and
$$r_i = \min_j(r(d'_{i}, d'_{\omega ij})).$$
$d'_{\omega ij}$ means the $j^{\text{th}}$ positive document for query $i$ and $r(d'_{i}, d'_{\omega ij})$ means the rank of $d'_{\omega ij}$ in the candidate set $d'_{i}$.
If we use 1-MRR@10 for the empirical risk function $\widehat{R}(\mathcal{T_{\lambda}}; \zeta)$, we could see that the risk function and its upper-bound plateaus after a certain $\lambda$ value as shown in Fig~\ref{fig:risk_function:truncate}, due to the threshold function in Eq~\eqref{eq:mrr_threshold}.
Therefore, naive certification will fail if the risk level is specified too low.

\subsection{Correction for Risk Level and Confidence}\label{sec:correction}
To this end, we propose a safe certification method, which will automatically correct the risk level $\alpha$ or the confidence $1-\delta$ if the risk level specified by the user is too low.
Given a specific $(\alpha, \delta)$ pair, the basic idea is that if the minimum of the upper confidence bound $\widehat{R}_{\delta}^+(\mathcal{T_{\lambda}};\zeta)$ over all the possible $\lambda$ value is bigger than the specified risk level $\alpha$, we will either replace the risk level with the best-possible minimal risk that we could achieve:
\begin{align}\label{eq:corrected_alpha}
  \alpha_c \doteq \inf_{\hat{\lambda}\in\Lambda}\widehat{R}_{\delta}^+(\mathcal{T_{\hat{\lambda}}};\zeta),  
\end{align}
where $\alpha_c$ is the calibrated risk level as shown in Fig.~\ref{fig:risk_function:correct_risk}, or increase $\delta$ to shrink the upper-confidence bound until $\delta=1$:
\begin{align}\label{eq:corrected_delta}
    \delta_c \doteq \inf\left\{\delta\in(0,1]: \inf_{\hat{\lambda}\in\Lambda}\widehat{R}^+_{\delta}(\mathcal{T_{\hat{\lambda}}};\zeta)\leq \alpha \right\},
\end{align}
where $\delta_c$ is the calibrated significance level as shown in Fig.~\ref{fig:risk_function:correct_conf}.
Therefore, the previous pruning function found in Eq.~\eqref{eq:lammbda_hat} does not necessarily hold for some specific $(\alpha, \delta)$ values in ranking, and we propose a new theorem for the corrected version of certification:

\begin{figure}[t]
\centering
\begin{subfigure}[t]{0.23\textwidth}
 \centering
\includegraphics[width=\textwidth]{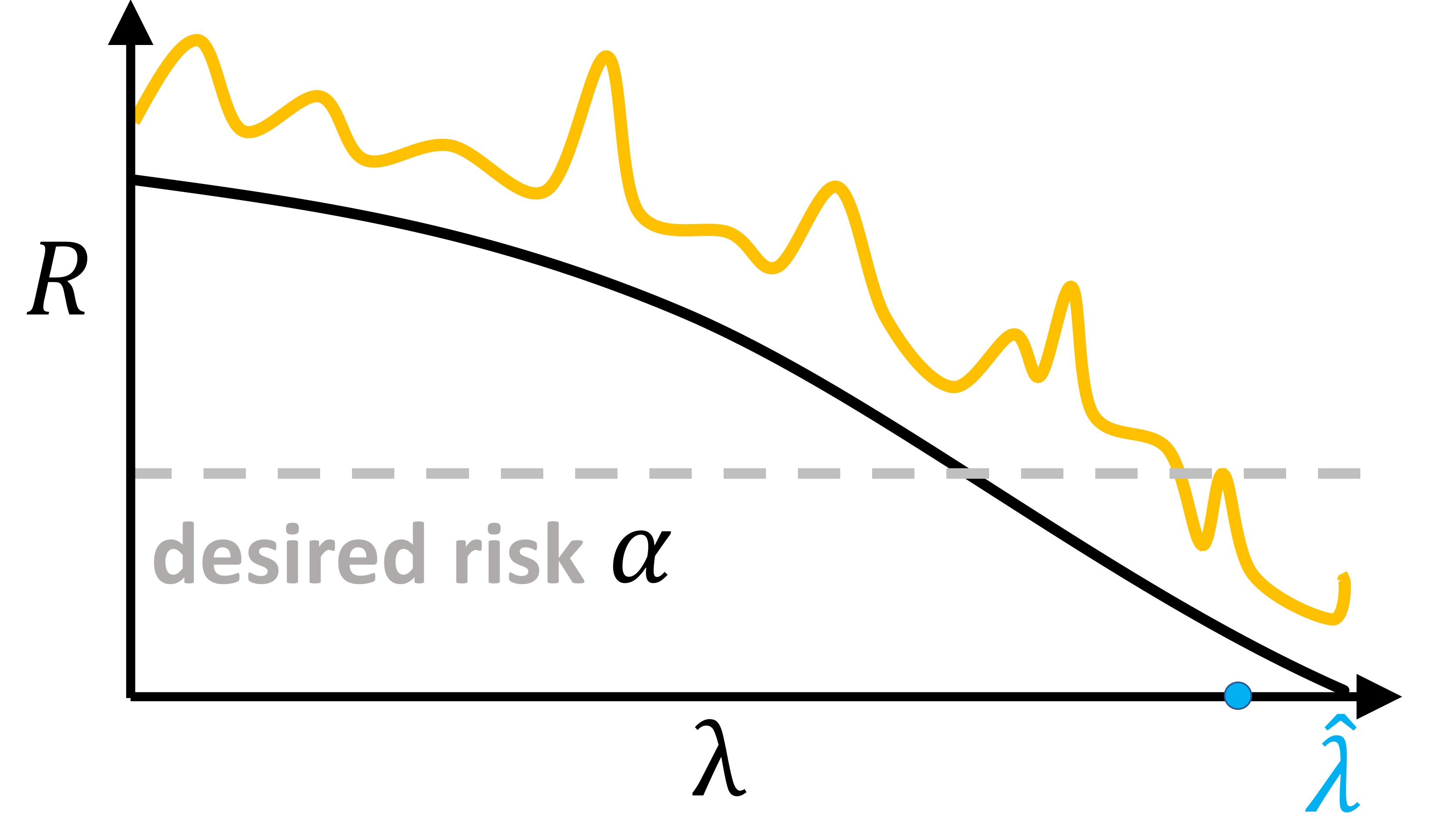}
\caption{Normal risk function and its upper bound.}
\label{fig:risk_function:naive}
\end{subfigure}
\begin{subfigure}[t]{0.23\textwidth}
 \centering
\includegraphics[width=\textwidth]{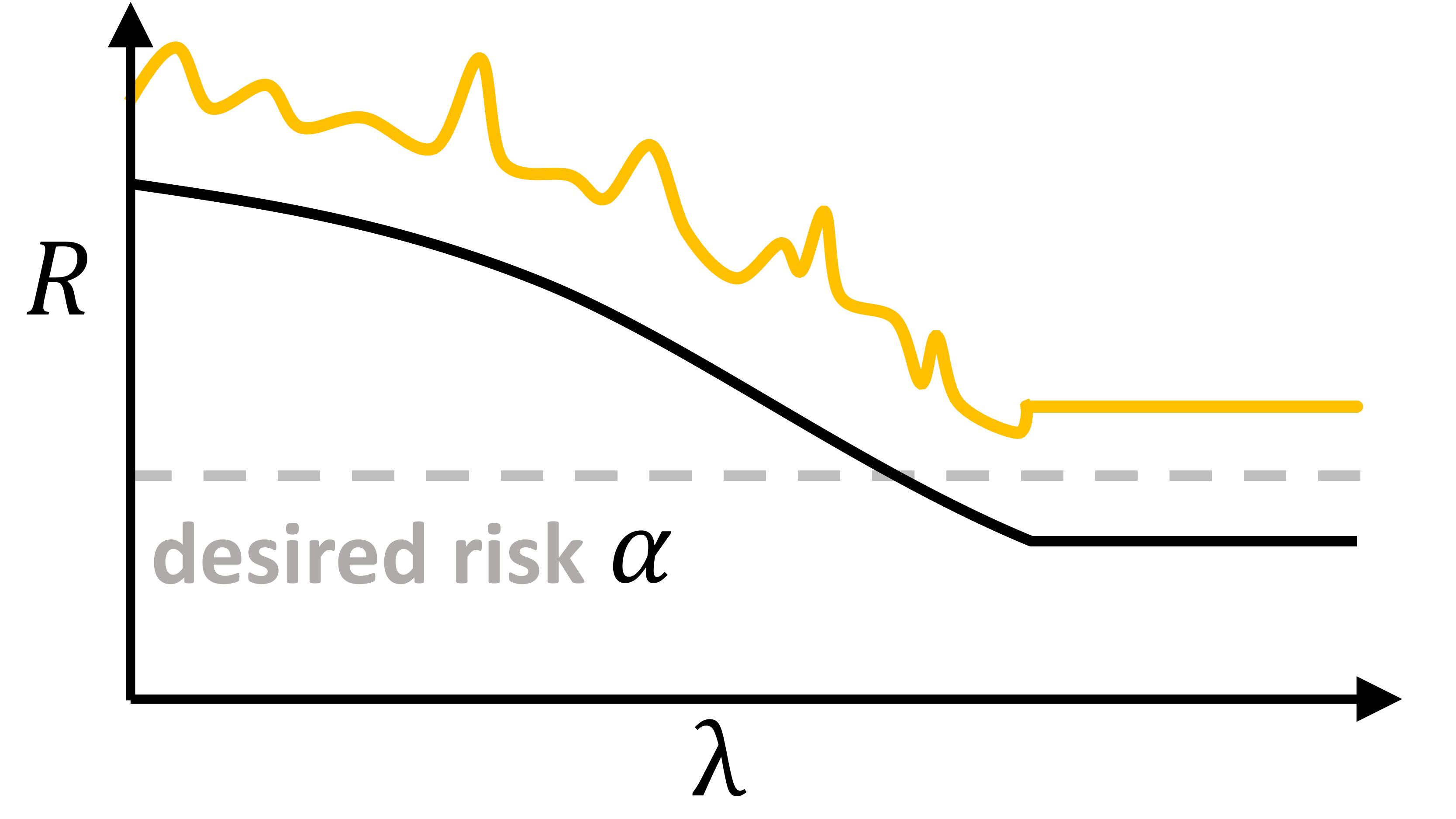}
\caption{Truncated risk function and its upper bound.}
\label{fig:risk_function:truncate}
\end{subfigure}
\begin{subfigure}[t]{0.23\textwidth}
 \centering
\includegraphics[width=\textwidth]{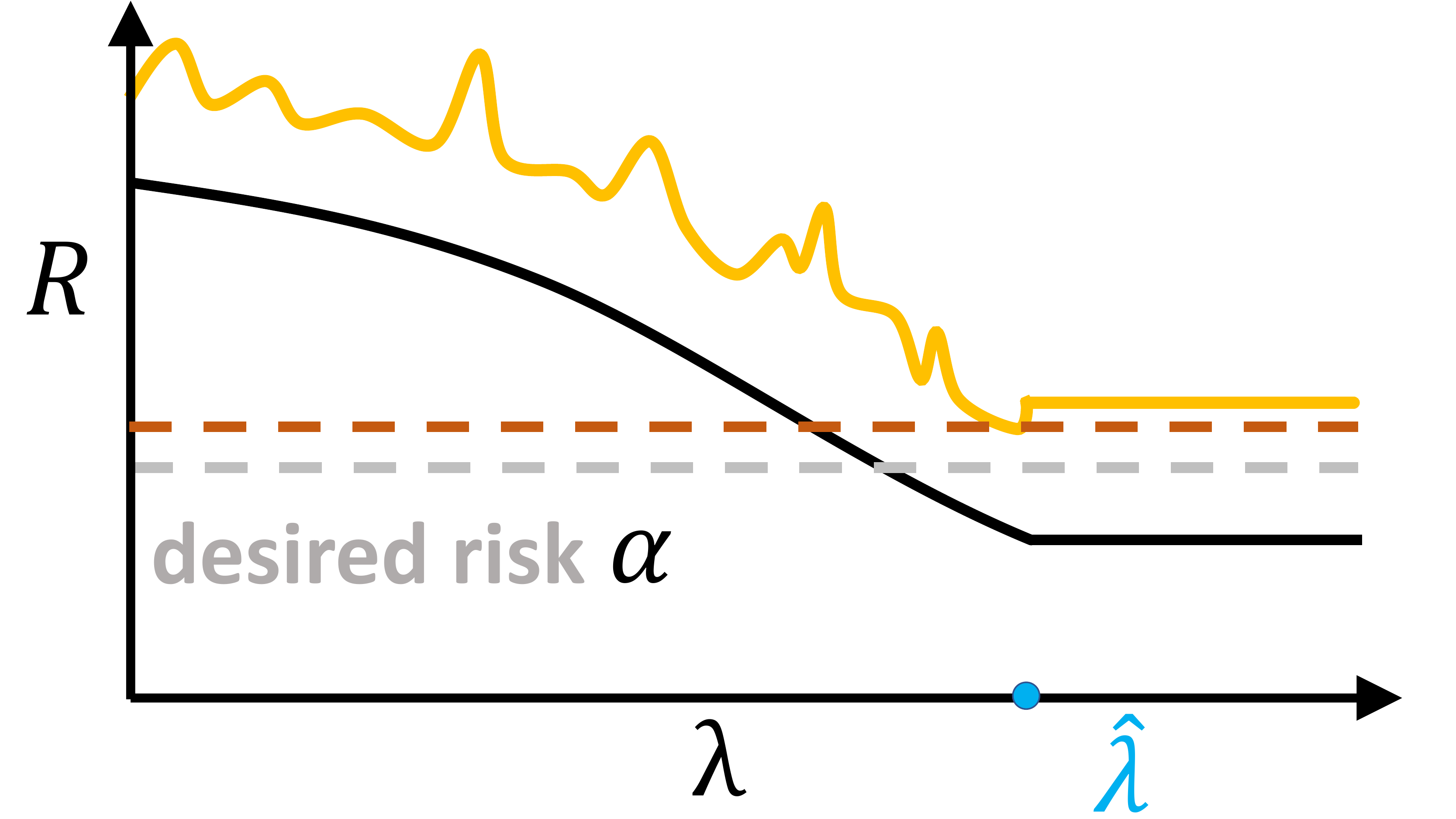}
\caption{Risk threshold correction.}
\label{fig:risk_function:correct_risk}
\end{subfigure}
\begin{subfigure}[t]{0.23\textwidth}
 \centering
\includegraphics[width=\textwidth]{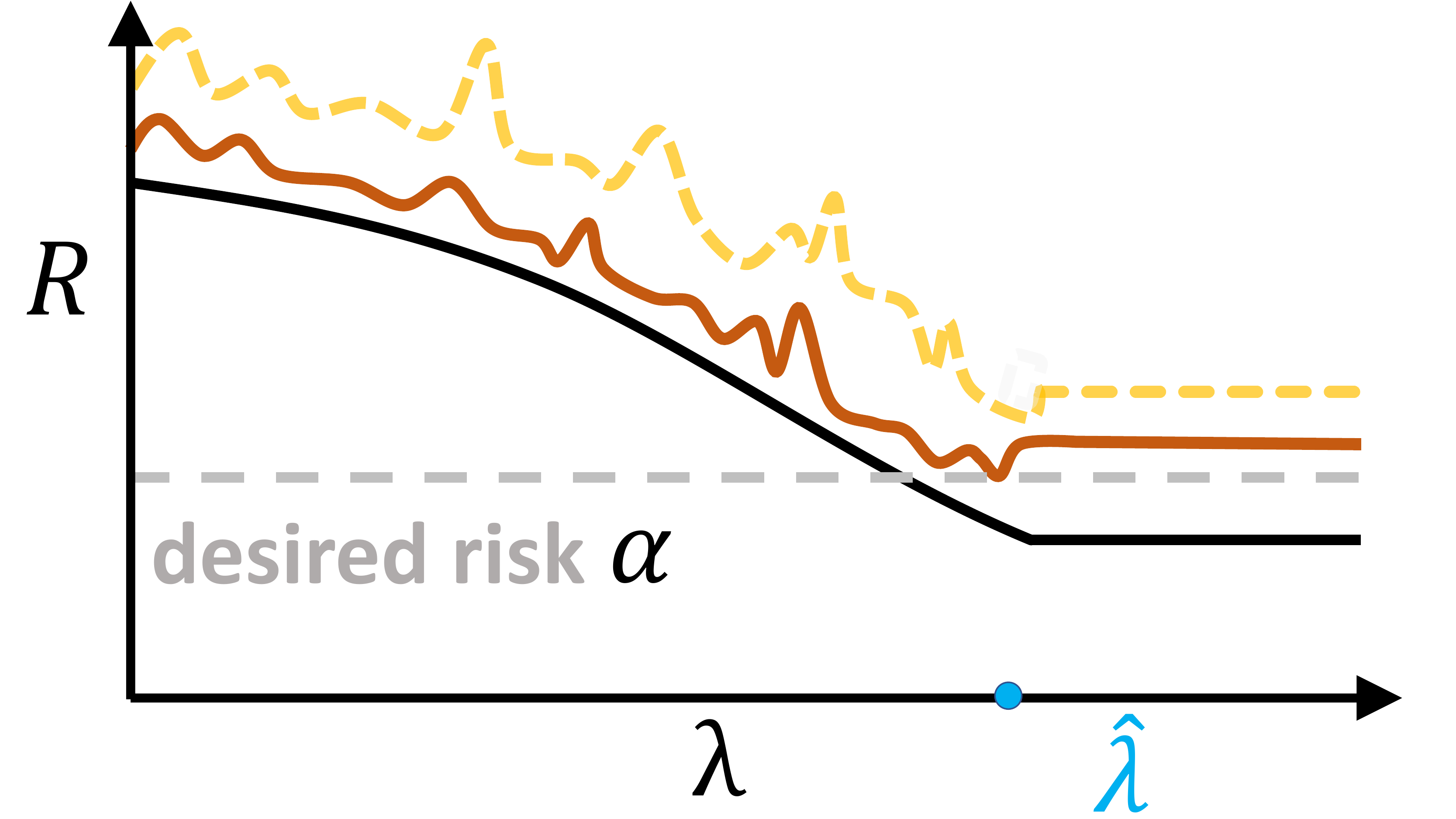}
\caption{Confidence correction.}
\label{fig:risk_function:correct_conf}
\end{subfigure}
\caption{(Corrected) Confidence region of the risk function. (a) shows the risk function for metrics like recall; (b) shows the risk function that is truncated in ranking. (c) and (d) show two different types of correction: Either setting the risk level to the minimum of the upper bound or increasing $\delta$ to shrink the upper bound.}
\label{fig:risk_function}
\end{figure}
\begin{theorem}[Correction of Certified Error Control]\label{thm:safeness}
In the setting of Section~\ref{sec:rcps_ranking}, assume that there exists $\alpha > 0$ and $\delta >0$ such that for every $\hat{\lambda}\in\Lambda$, $\widehat{R}^+_{\delta}(\mathcal{T_{\hat{\lambda}}};\zeta) > \alpha$. In this case, $\mathcal{T}_{\hat{\lambda}}$ is no longer a pruning function with certified error control.
Instead, with the corrected risk level $\alpha_c$ and confidence $\delta_c$ in Eq~\eqref{eq:corrected_alpha} and~\eqref{eq:corrected_delta}, there exists $\hat{\lambda}\in\Lambda$ such that,
$$\Pr(R(\mathcal{T}_{\hat{\lambda}};\zeta)\leq \alpha_c) \geq 1-\delta,$$
or
$$\Pr(R(\mathcal{T}_{\hat{\lambda}};\zeta)\leq \alpha) \geq 1-\delta_c.$$
In both cases, $\mathcal{T_{\hat{\lambda}}}$ is a pruning function with certified error control.
\end{theorem}
Proofs are provided in Appendix~\ref{sec:appendix:proofs}.
In addition, we provide an algorithmic implementation in Appendix~\ref{sec:appendix:algorithm} for readers' further reference.

\section{Experimental Setup}
\subsection{Datasets}
In this paper, we evaluate our method on the following datasets. MS MARCO Passage v1~\citep{nguyen2016ms} contains 8.8M English passages of average length around 55 tokens, which is a standard retrieval benchmark for comparing in-domain results.
Quora Duplicate Questions\footnote{https://quoradata.quora.com/First-Quora-Dataset-Release-Question-Pairs}~\citep{thakur2021beir} contains 522K passages of average length around 11 tokens and mostly consists of duplicate, entity questions which were found challenging for out-of-domain generalization of neural retrievers like DPR~\citep{DBLP:journals/corr/abs-2109-08535}.
The above datasets label documents with binary judgements for each query and therefore MRR@10 is often used as the evaluation metric.
Other datasets such as TREC DL 2019~\citep{DBLP:conf/trec/2019} use metrics like nDCG, but such densely labelled data are very scarce and therefore they are not suitable for finite-sample calibration, which we leave for future work.

\begin{table}[t!]
\centering
\begin{small}
\begin{tabular}{l@{\hskip 1.2in}c@{\hskip 0.05in}c@{\hskip 0.05in}}
\toprule
\textbf{Retriever}&MARCO &Quora\\
\midrule
BM25 &0.185&0.781\\
DPR &0.311&0.434\\
UniCOIL &0.348&0.659\\
\bottomrule
  \end{tabular}
\end{small}
  \caption{In-domain (MS MARCO Passage v1) and out-of-domain (Quora) first-stage retrieval test MRR@10 score averaged over 100 random dev/test splits.
}
    \label{tbl:marco_retrieve}
\end{table}

\begin{table}[t!]
\centering
\begin{small}
\begin{tabular}{l@{\hskip 0.3in}c@{\hskip 0.05in}c@{\hskip 0.05in}}
\toprule
\textbf{Retriever}&MARCO &Quora\\
\midrule
BM25+MonoBERT &0.369&0.840\\
DPR+MonoBERT &0.378&0.728\\
UniCOIL+MonoBERT &0.383&0.832\\
\midrule
BM25+MonoELECTRA &0.399&0.823\\
DPR+MonoELECTRA &0.415&0.653\\
UniCOIL+MonoELECTRA &0.415&0.817\\
\bottomrule
  \end{tabular}
\end{small}

  \caption{In-domain (MS MARCO Passage v1) and out-of-domain (Quora) second-stage reranking (with evidence fusion) test MRR@10 score averaged over 100 random dev/test splits.
}
    \label{tbl:marco_rerank}
\end{table}

\begin{table*}[!t]
\centering
% \resizebox{16cm}{!}{
\begin{subtable}{.23\linewidth}
\hskip-4cm\resizebox{7.5cm}{!}{
\begin{tabular}{l@{\hskip 0.2in}c@{\hskip 0.1in}c@{\hskip 0.1in}c@{\hskip 0.1in}c@{\hskip 0.1in}}
\toprule
\textbf{Methods}&$\text{MRR@10}$&Confidence &Coverage &Size\\
\midrule
% monoBERT: 0.00673352436 sec / (query, top-1 doc)
\multicolumn{5}{l}{\small{\textit{BM25 + MonoBERT (required MRR@10=0.350)}}}\\
\quad CEC &0.359&0.870&\textbf{0.870}&451\\
\quad EST &0.350&-&0.540&139\\
\quad ERT &0.350&-&0.510&149\\
\hline
\multicolumn{5}{l}{\small{\textit{UniCOIL + MonoBERT (required MRR@10=0.350)}}}\\
\quad CEC &0.360&0.900&\textbf{0.900} &15 \\
\quad EST &0.350&-&0.480&9\\
\quad ERT &0.352&-&0.460&7\\
\hline
\multicolumn{5}{l}{\small{\textit{DPR + MonoBERT (required MRR@10=0.350)}}}\\
\quad CEC &0.360&0.900&\textbf{0.900} &22  \\
\quad EST&0.350&-&0.500&11\\
\quad ERT &0.353&-&0.730&11\\
\midrule
% monoELECTRA: 0.0032951289 sec / (query, top-1 doc)
\multicolumn{5}{l}{\small{\textit{BM25 + MonoELECTRA (required MRR@10=0.380)}}}\\
\quad CEC&0.389&0.894&\textbf{0.910}&602 \\
\quad EST&0.381&-&0.580&280\\
\quad ERT &0.381&-&0.550&246\\
\hline
\multicolumn{5}{l}{\small{\textit{UniCOIL + MonoELECTRA (required MRR@10=0.380)}}}\\
\quad CEC &0.390&0.900 &\textbf{0.910} &18 \\
\quad EST &0.380&-&0.520&13\\
\quad ERT &0.383&-&0.750&9\\
\hline
\multicolumn{5}{l}{\small{\textit{DPR + MonoELECTRA (required MRR@10=0.380)}}}\\
\quad CEC &0.389& 0.900&\textbf{0.900} &27 \\
\quad EST &0.381&-&0.580&16\\
\quad ERT &0.382&-&0.580&17\\
\bottomrule
  \end{tabular}
}
\end{subtable}
\begin{subtable}{.23\linewidth}
\resizebox{7.5cm}{!}{
\begin{tabular}{l@{\hskip 0.2in}c@{\hskip 0.1in}c@{\hskip 0.1in}c@{\hskip 0.1in}c@{\hskip 0.1in}}
\toprule
\textbf{Methods}&$\text{MRR@10}$&Confidence &Coverage &Size\\
\midrule
\multicolumn{5}{l}{\small{\textit{BM25 + MonoBERT (required MRR@10=0.780)}}}\\
\quad CEC &0.789 &0.900&\textbf{0.910}&2\\
\quad EST &0.780&-&0.510&2\\
\quad ERT &0.780&-&0.600&3\\
\hline
\multicolumn{5}{l}{\small{\textit{UniCOIL + MonoBERT (required MRR@10=0.780)}}}\\
\quad CEC &0.790&0.900&\textbf{0.900} &16 \\
\quad EST &0.780&-&0.560&13\\
\quad ERT &0.782&-&0.640&14\\
\hline
\multicolumn{5}{l}{\small{\textit{DPR + MonoBERT (required MRR@10=0.620)}}}\\
\quad CEC &0.632&0.900&\textbf{0.940} &40  \\
\quad EST&0.620&-&0.500&30\\
\quad ERT &0.621&-&0.490&29\\
\midrule
\multicolumn{5}{l}{\small{\textit{BM25 + MonoELECTRA (required MRR@10=0.780)}}}\\
\quad CEC&0.790&0.900&\textbf{0.900}&3 \\
\quad EST&0.780&-&0.460&3\\
\quad ERT &0.780&-&0.560&2\\
\hline
\multicolumn{5}{l}{\small{\textit{UniCOIL + MonoELECTRA (required MRR@10=0.780)}}}\\
\quad CEC &0.790&0.900 &\textbf{0.910} &3 \\
\quad EST &0.785&-&0.540&3\\
\quad ERT &0.782&-&0.600&3\\
\hline
\multicolumn{5}{l}{\small{\textit{DPR + MonoELECTRA (required MRR@10=0.620)}}}\\
\quad CEC &0.634& 0.900&\textbf{0.950} &282 \\
\quad EST &0.620&-&0.530&155\\
\quad ERT &0.620&-&0.540&149\\
\bottomrule
  \end{tabular}
}
\end{subtable}

% }
\caption{In-domain results on MS MARCO Passage v1 (left) and out-of-domain results on Quora (right). CEC: Certified error control. EST: Empirical score threshold. ERT: Empirical rank threshold.  Confidence: $1-\delta_c$ as in Eq.~\eqref{eq:corrected_delta}. Coverage: Proportion of 100 trial runs that satisfy the risk constraint. Size: Average candidate set size out of 1,000. 
See Section~\ref{sec:results:coverage} for details.
}
    \label{tbl:marco_coverage}
\end{table*}

\subsection{Retrievers and Rerankers}
For retrievers, we use DPR (dense retriever), BM25 (static lexical retriever), and UniCOIL (contextualized lexical retriever).
For rerankers, we use cross-encoder models like MonoELECTRA and MonoBERT~\citep{Nogueira2019MultiStageDR}.
Although some of the models are no longer state of the art, they remain competitive and have been widely adopted by other members of the community as points of reference.
For the pipeline, we use the retriever (e.g., DPR) to retrieve the top-1,000 candidates from the corpus, and then use the reranker (e.g., MonoELECTRA) to rerank the retrieved candidate sets.
We believe the above choices cover the basic types of modern two-stage ranking systems; our approach is model agnostic and can be easily applied to other models as well.
For implementation, we use off-the-shelf pre-trained models from Pyserini~\citep{lin2021pyserini} and Caprelous~\cite{yates2020caprelous}.

Finally, for rerankers based on neural networks, we need to consider both in-domain and out-of-domain situations, as it is possible that the reranker overfits to certain domains and has worse out-of-domain performance than the retriever.
To solve this, we linearly interpolates the score of the retriever $\phi$ and the reranker $\zeta$ for each $(q,d)$ pair during reranking:
\begin{align}
    s_f(q,d) = &\beta\cdot\phi(\eta_q (q), \eta_d (d)) +\nonumber\\ &(1-\beta)\cdot\zeta (\text{concat}(q, d))\nonumber,
\end{align}
which is known as evidence fusion~\citep{ma2022dprrepro}.
The weight $\beta\in[0,1]$ is searched on the calibration set for the best MRR@10 score, such that the fusion model will consistently yield better ranking results than both $\zeta$ and $\phi$.
Tbl.~\ref{tbl:marco_retrieve} and~\ref{tbl:marco_rerank} show the retrieval and reranking performance with evidence fusion under both in-domain and out-of-domain settings.

\begin{figure}[t!]
    \centering
    \includegraphics[width=0.45\textwidth]{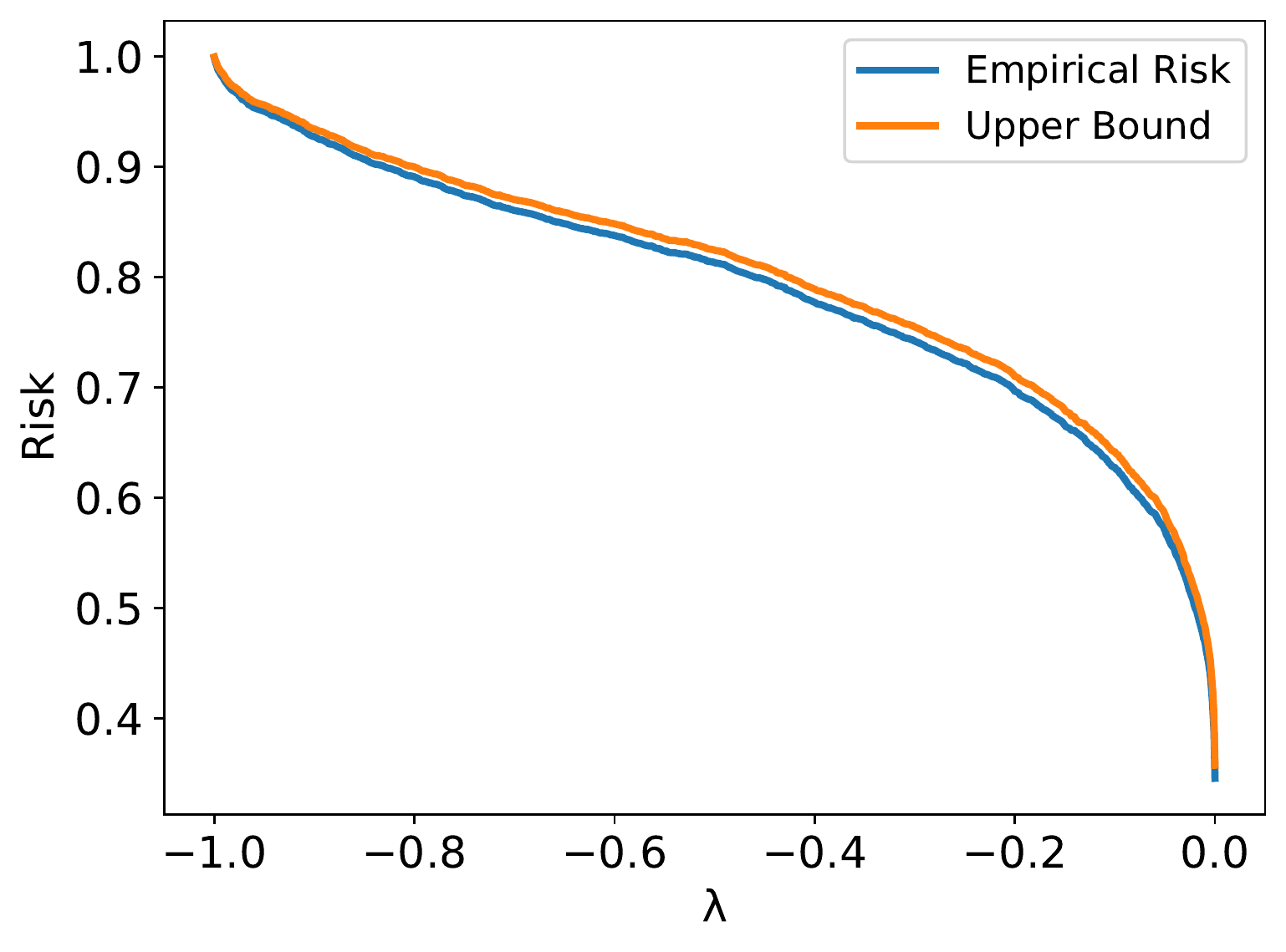}
    \caption{Empirical risk function and its upper bound on one calibration/test data split.}
    \label{fig:risk_ucb}
\end{figure}

\subsection{Baselines}
We modify two \textit{empirical error control} methods from~\citet{Cambazoglu2010EarlyEO} as the baselines:

\smallskip
\noindent\textbf{Empirical Score Threshold (EST).} A score threshold on the calibration set such that the pruned MRR@10 just meets the required score.

\smallskip \noindent\textbf{Empirical Rank Threshold (ERT).} Similar to EST, but we tune a threshold on the rank of the documents to prune the candidate set instead.

\section{Results}
\subsection{Risk Function and Upper Bound}
Fig.~\ref{fig:risk_ucb} shows the empirical risk function and its WSR upper bound of the pruning function $\mathcal{T}_{\lambda}$ on Quora using the DPR + MonoELECTRA ranking system.
We could see that the empirical risk function is a monotone function and the minimum of the risk is greater than 0, which is consistent with the assumptions we made in Section.~\ref{sec:methods}.
In addition, we could see that the bound is very tight, providing a good estimation of the true risk.

\subsection{In-Domain Results}\label{sec:results:coverage}
Empirically, if we set the risk threshold $\alpha=0.62$ and confidence level $1-\delta=0.9$, then there should be at least $90\%$ of independent runs (i.e., coverage) that the MRR@10 score is greater than 0.38 (i.e., $1-\alpha$).
To verify this, we mix the test set and dev set and then randomly sample a calibration set of size 5,000 and a test set of size 6,980 for 100 trials.

Tbl.~\ref{tbl:marco_coverage} (left) shows the MRR@10 score, corrected confidence, empirical coverage, and average candidate set size on MS MARCO Passage v1.
We choose different performance thresholds for different ranking systems such that the final reranking score is dropped around $1\sim2\%$, which is a very typical choice in applications.
We use $\delta=0.1$ for all experiments, but it could be corrected if the risk threshold is unable to satisfy.
For example, for BM25 + MonoBERT, the confidence is corrected from $0.90$ to $0.87$, which is more consistent with the empirical coverage.

We could see our method achieves the required risk constraint with the required coverage (i.e., confidence) for multiple ranking systems, while the average candidate set size is also drastically reduced from 1,000 to less than 50.
In comparison, although being able to obtain a smaller candidate set size, both empirical error control methods do not achieve the expected coverage.
Despite that we could choose other thresholds to achieve better coverage, it is unclear how much accuracy should be sacrificed in order to achieve the required coverage.

\subsection{Out-of-Domain Results}
We also test our method on Quora under an out-of-domain setting, where we use the retrievers and rerankers trained on MS MARCO Passage v1 as the prediction models.
We can see from Tbl.~\ref{tbl:marco_retrieve} and~\ref{tbl:marco_rerank} that the out-of-domain retrieval and reranking results are drastically different from the MS MARCO dataset, where BM25 outperforms the other neural retrievers.
This is because the Quora dataset mostly consists of duplicate, entity-based questions, which is naturally biased towards static lexical retrievers.

Similar to the in-domain experiments, we calibrate the pruning function on the calibration set with 5,000 data points and test it on a test set with 10,000 data points over 100 trial runs.
Tbl.~\ref{tbl:marco_coverage} (right) shows the MRR@10 scores of different ranking systems.
However, unlike the in-domain setting, the ranking performance of the first stage retrievers varies a lot under the out-of-domain setting and it is hard to align the pruning results for intuitive comparison.
Therefore, we set different MRR@10 thresholds such that the performance drop is around $1\sim10\%$ to align the results of different ranking systems.
The results are similar to the in-domain experiments, where the certified error control method manages to provide the performance guarantee while the empirical method fails to.
This is consistent with our statements that our method does not make assumptions about data distribution and prediction models, as long as the data of the calibration and test set are exchangeable.

\begin{figure}[t!]
    \centering
    \includegraphics[width=0.45\textwidth]{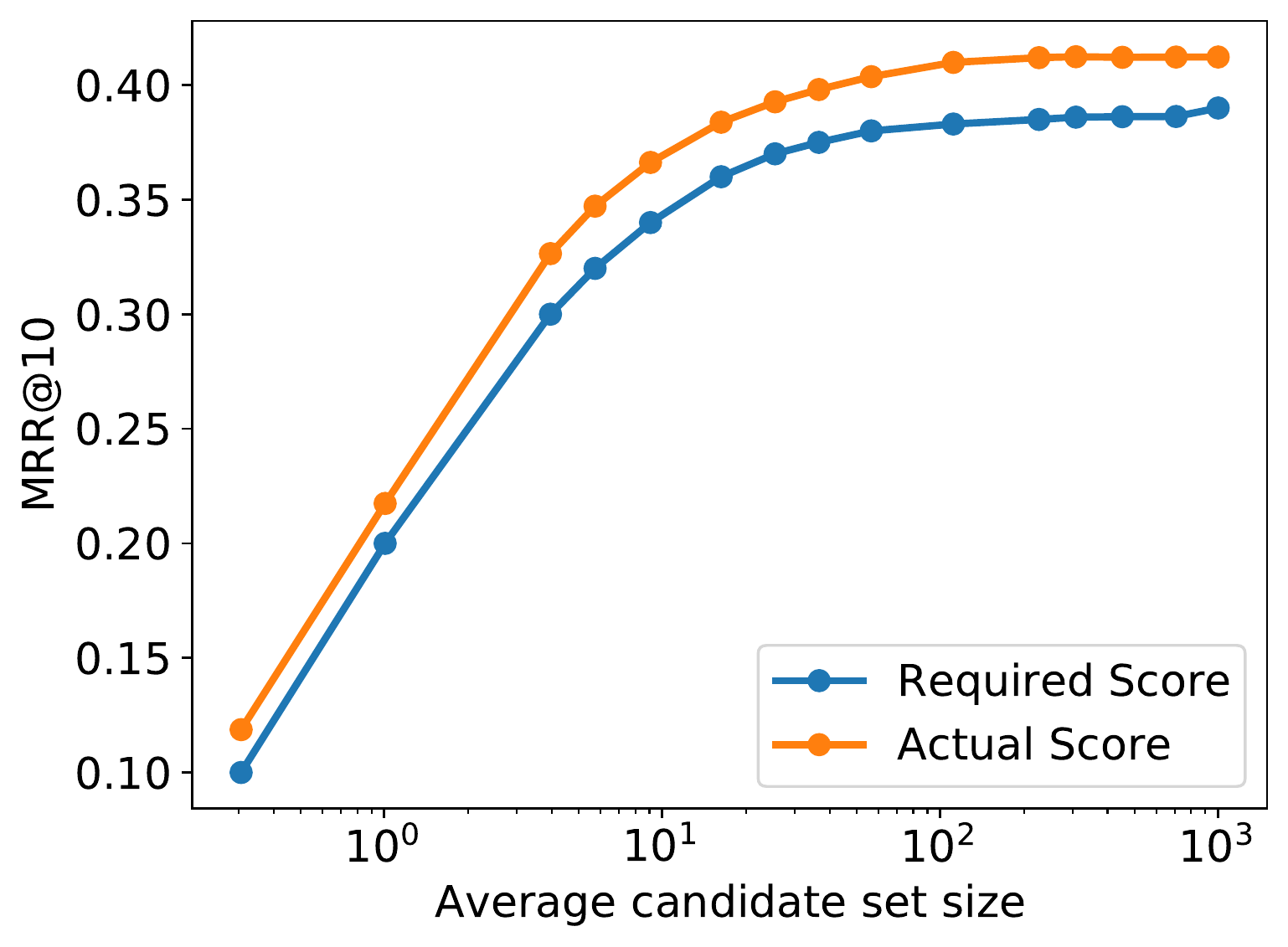}
    \caption{Tradeoffs between candidate set size and MRR@10 score on MS MARCO Passage v1 based on DPR + MonoELECTRA ranking system. The empirical coverage for each point is greater than 90\%.}
    \label{fig:trade_off}
\end{figure}

\subsection{Efficiency-Accuracy Tradeoffs with Certified Error Control}\label{sec:results:tradeoffs}
In this section, we investigate the guarantee of the overall tradeoffs between efficiency and accuracy.
Fig.~\ref{fig:trade_off} illustrates the efficiency-accuracy tradeoff results on MS MARCO Passage v1 using DPR + MonoELECTRA.
Similarly, we set the confidence level $1-\delta$ to be $0.9$.
Our method (blue line) achieves the best MRR@10 score at around $20\%$ of the original top-1,000 candidate set size, which is a very good tradeoff between accuracy and efficiency.
In addition, the MRR@10 score of our method (blue line) is higher than the specified score threshold (orange line) with at least 90\% coverage, which further verifies the guarantee claims we made about our method.
We could see that our method achieves a good tradeoff while satisfying different values of the risk level $\alpha$.

\begin{figure}[t!]
    \centering
    \includegraphics[width=0.45\textwidth]{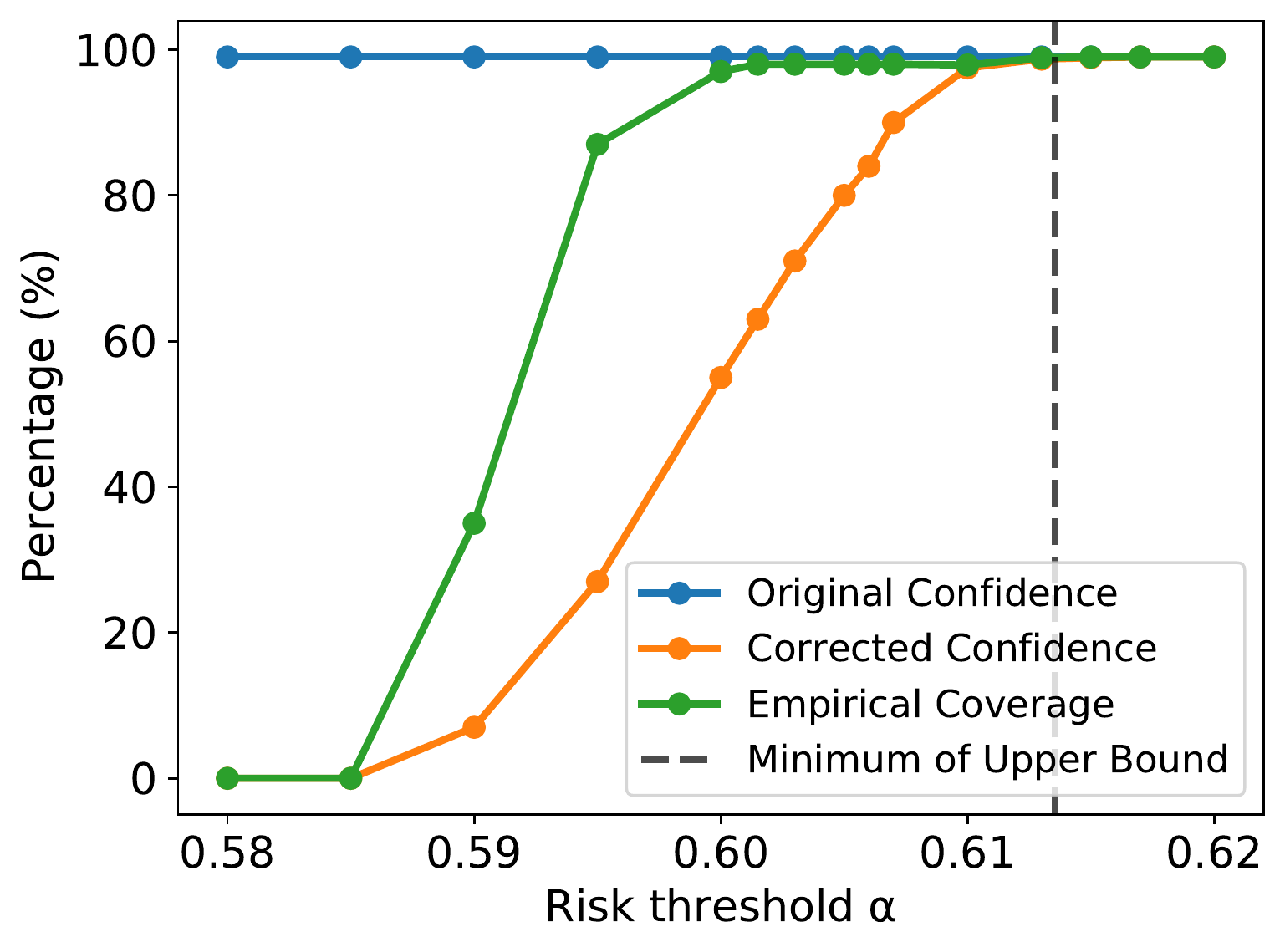}
    \caption{Confidence correction on MS MARCO Passage v1 using DPR + MonoELECTRA. The x-axis represents different risk thresholds (i.e., $\alpha$) and the y-axis represents the percentage.}
    \label{fig:conf_correction}
\end{figure}

\subsection{Confidence-Risk Correction}\label{sec:results:correction}
In Section~\ref{sec:correction}, we mentioned that it might be impossible to achieve the risk threshold if specified too low in ranking tasks.
Our solution approaches this problem by correcting either the risk threshold $\alpha$ or the significance level $\delta$ as shown in Fig.~\ref{fig:risk_function:correct_risk} and~\ref{fig:risk_function:correct_conf}.
In practice, the risk threshold correction is rather straightforward: if the risk is lower than the minimal upper bound over all $\lambda$, we just reset the risk threshold to the minimal upper bound as shown in Eq.~\eqref{eq:corrected_alpha}.
For the confidence correction, we need to fix the risk threshold and shrink the upper bound by increasing the significance $\delta$ until the confidence ($1-\delta$) is 0 as shown in Eq.~\eqref{eq:corrected_delta}.
Fig.~\ref{fig:conf_correction} shows the confidence correction of our method.
We use the DPR+MonoELECTRA ranking system whose best reranking MRR@10 score is $0.41$ on the dev small set, meaning that the minimal risk is around $0.59$, and the minimal UCB of the risk function is around $0.61$ (the vertical line).
From right to left, we could see that the confidence does not change too much until the risk passes $0.6137$.
As the risk threshold decreases, the confidence (the orange line) also gradually decreases which is consistent with the empirical coverage (the green line).

\section{Conclusion}
We present a theoretically principled method which provides certified error control of candidate set pruning for two-stage ranking systems, allowing users to accurately control a customized loss under the desired threshold with high probability.
We further propose to correct the risk threshold or confidence level if the desired risk cannot be achieved given the ranking system.
Experiments performed under in-domain (MS MARCO Passage v1) and out-of-domain (Quora) settings show that our method provides a consistent performance guarantee to the candidate set pruning across multiple ranking systems.

\section*{Acknowledgements}
This research was supported in part by the Canada First Research Excellence Fund and the Natural Sciences and Engineering Research Council (NSERC) of Canada; computational resources were provided by Compute Canada.
\bibliography{anthology,custom}
\bibliographystyle{acl_natbib}

\clearpage
\appendix

\section{Appendix}
\label{sec:appendix}
\subsection{Proofs}\label{sec:appendix:proofs}
\noindent\textit{Proof of Theorem \ref{thm:safeness}.}
We first prove that if there exists $\alpha > 0$ and $\delta >0$ such that for every $\hat{\lambda}\in\Lambda$, $\widehat{R}^+_{\delta}(\mathcal{T_{\hat{\lambda}}};\zeta) > \alpha$, then $\mathcal{T}_{\hat{\lambda}}$ will no longer be a pruning function certified error control.
Suppose there exists an $\hat{\lambda}$ such that $\Pr(R(\mathcal{T}_{\hat{\lambda}};\zeta)\leq \alpha) \geq 1-\delta$, by the coverage property in Eq.~\eqref{eq:ucb} we know that $\widehat{R}^+_{\delta}(\mathcal{T_{\hat{\lambda}}};\zeta) \leq \alpha$. Contradiction.

Next, we prove that for $(\alpha_c, \delta)$ in Eq.~\eqref{eq:corrected_alpha}, $\mathcal{T}_{\hat{\lambda}}$ has certified error control.
By definition in Eq.~\eqref{eq:corrected_alpha}, we know that there exists an $\hat{\lambda}$ such that $\widehat{R}^+_{\delta}(\mathcal{T_{\hat{\lambda}}};\zeta) = \alpha_c$, and by the coverage property in Eq.~\eqref{eq:ucb}, we have $\Pr(R(\mathcal{T}_{\hat{\lambda}};\zeta)\leq \alpha_c) \geq 1-\delta$. Done.

Finally, we prove that for $(\alpha_c, \delta)$  in Eq.~\eqref{eq:corrected_delta}, $\mathcal{T}_{\hat{\lambda}}$ has certified error control.
By definition in Eq.~\eqref{eq:corrected_delta}, we know that there exists an $\hat{\lambda}$ such that $\widehat{R}^+_{\delta_c}(\mathcal{T_{\hat{\lambda}}};\zeta) = \alpha$, and by the coverage property in Eq.~\eqref{eq:ucb}, we have $\Pr(R(\mathcal{T}_{\hat{\lambda}})\leq \alpha;\zeta) \geq 1-\delta_c$.
$\square$

\medskip
\begin{theorem}[Validity of Certified Error Control ]~\citep{Bates2021DistributionFreeRP}\label{thm:rcps}
Let $\{P'_i, {D'_\omega}_i\}_{i=1}^m$ be an i.i.d. sample and $L(P', {D'_\omega}; \zeta)$ is monotone w.r.t. $\lambda$ as in Eq~\eqref{eq:monotonicity}. 
Let $\{\mathcal{T}_{\lambda}\}_{\lambda\in\Lambda}$ be a collection of pruning function satisfying the nesting property in Eq~\eqref{eq:set_predictor}. 
Suppose Eq~\eqref{eq:ucb} holds pointwise for each $\lambda$, and that $R(\mathcal{T}_\lambda;\zeta)$ is continuous. Then for $\hat{\lambda}$ chosen as in Eq~\eqref{eq:lammbda_hat}, we have
$$\Pr(R(\mathcal{T}_{\hat{\lambda}}; \zeta)\leq \alpha) \geq 1-\delta.$$
That is, $\mathcal{T}_{\hat{\lambda}}$ is a pruning function with certified error control.
\end{theorem}

\medskip \noindent
\textit{Proof of Theorem \ref{thm:rcps}.}
Our proof follows the framework in
\citep{Bates2021DistributionFreeRP}.
Consider the smallest $\lambda$ that controls the risk:
\begin{align}\label{eq:lammbda_hat}
\lambda^* \doteq \text{inf}\left\{ \lambda \in \Lambda:R(\mathcal{T_{\lambda}};\zeta)\leq\alpha  \right\}.\nonumber
\end{align}
Suppose $R(\mathcal{T}_{\hat{\lambda}};\zeta) > \alpha$. By the definition of $\lambda^*$ and the monotonicity and continuity of $R(\cdot;\zeta)$, this implies $\lambda^*<\hat{\lambda}$.
By the definition of $\hat{\lambda}$, this further implies that $\widehat{R}^+_{\delta}(\mathcal{T}_{\lambda^*}) < \alpha$. But since $R(\mathcal{T}_{\lambda^*};\zeta) = \alpha$ (by continuity) and by
the coverage property in Eq.~\eqref{eq:ucb}, this happens with probability at most $\delta$. 
$\square$

\subsection{Waudby-Smith–Ramdas Bound}\label{sec:appendix:wsr}
\citet{Bates2021DistributionFreeRP}~provide a one-sided variant of the Waudby-Smith–Ramdas (WSR) bound \citep{WaudbySmith2020VarianceadaptiveCS,Bates2021DistributionFreeRP}:
\begin{proposition}[Waudby-Smith–Ramdas bound]
Let $L_i(\lambda) = L(D'_{\omega}, T_{\lambda}(D');\zeta)$, and
\begin{align}
    \hat{\mu}_i(\lambda) =& \frac{\frac{1}{2}+\sum_{j=1}^{i}L_j(\lambda)}{1+i},\nonumber\\ 
    \hat{\sigma}_i^2(\lambda)=&\frac{\frac{1}{4}+\sum_{j=1}^{i}(L_j(\lambda)-\hat{\mu}_i(\lambda))^2}{1+i},\nonumber\\
    \nu_i(\lambda) =&\min\left\{1, \sqrt{\frac{2\log(1/\delta)}{n\hat{\sigma}_i^2(\lambda)}} \right\}. \nonumber
\end{align}
Further let
$$\mathcal{K}_i(R;\lambda)=\prod_{j=1}^{i}\{1 - \nu_j(\lambda)(L_j(\lambda) - R)\} ,$$
and
$$\widehat{R}_{\delta}^+(\mathcal{T}_\lambda) = \inf\left\{
R \geq 0: \max_i \mathcal{K}_i(R; \lambda) > \frac{1}{\delta}
\right\}
.$$
Then $\widehat{R}_{\delta}^+(\mathcal{T}_\lambda)$ is a $(1 - \delta)$ upper confidence bound for $R(\lambda)$.
\end{proposition}

\medskip \noindent
The proofs are basically a restatement of the Theorem 4 in~\citet{WaudbySmith2020VarianceadaptiveCS} and Proposition 5 in~\citet{Bates2021DistributionFreeRP}.

\subsection{Algorithmic Implementation}\label{sec:appendix:algorithm}
Alg.~\ref{alg:rcps_calib} provides a detailed implementation of the certified error control method for relevance ranking using the form of pseudo-code, where we use MRR@10 as the metric. 
The algorithm takes the ranking system and calibration set as the inputs and returns the set predictor, the corrected risk, and corrected confidence.

\begin{algorithm*}[!t]
\caption{Calibration procedure.}\label{alg:rcps_calib}
\KwParameter{Risk Level $\alpha$, Confidence $1-\delta$}
\KwModel{Retriever $\{\eta_Q, \eta_D\}$, Reranker $\zeta$}
\KwData{Calibration Set $\{q_i
, d'_i, {d'_\omega}_i\}_{i=1}^m$}
\KwMetric{MRR@10}
\KwResult{$\hat{\lambda}$, $\alpha_c$, $1-\delta_c$}
\texttt{/*Retrieval Prediction*/}\\
$S'_v \gets \emptyset, D'_v \gets \emptyset$\\
\For{$i\gets1$ \KwTo $m$}{
$u_i \gets \eta_Q(q_i)$\\
$S_v \gets \emptyset,D_v \gets \emptyset$\\
\For{$j\gets1$ \KwTo $k$}{
    $v_{ij} \gets \eta_D(d'_{ij}),s_v = u_i^Tv_{ij}$\\
    $S_v\gets S_v\cup\{s_v\},D_v\gets D_v\cup\{d'_{ij}\}$\\
    }
    Sort $D_v$ and $S_v$ in desc. order of $S_v$\\
    $S'_v\gets S'_v\cup S_v,D'_v\gets D'_v\cup D_v$\\
}
$S'_v\gets$ Platt-Scaling$(S'_v)$\\ 
\texttt{/*Reranking and Compute Upper Bound*/}\\
$\widehat{R}^+ \gets \emptyset, L' \gets \emptyset, \Lambda \gets \emptyset$\\

\For{$\lambda\gets1$ \KwTo $0$ \KwBy $-10^{-5}$}{
 $P'\gets D'_{v (S'_v\geq \lambda)}, L\gets\emptyset$\\
\For{$i\gets1$ \KwTo $m$}{
$S_r\gets\emptyset,P_r\gets\emptyset$\\
\For{$p$ \KwIn $P'_i$}{
$s_r = \beta\cdot\phi(\eta_q (q_i), \eta_d (p)) +(1-\beta)\cdot\zeta (\text{concat}(q, p))$\\
$S_r\gets S_r\cup\{s_r\}, P_r\gets P_r\cup\{p\}$\\
Sort $P_r$ in desc. order of $S_r$\\
}
$L \gets L \cup \{1-\text{MRR@10}(P_r, d'_{\omega i})\}$\\
}
$\widehat{R}^+ \gets \widehat{R}^+ \cup \{\text{WSR}(L, \delta)\}$\\
$\Lambda \gets \Lambda \cup \{\lambda\}, L' \gets L' \cup L$
}
\texttt{/*Compute Lambda*/}\\
\eIf{$\min(\widehat{R}^+) \leq \alpha$}{
  \For{$\hat{\lambda}, R$ \KwIn $\Lambda, \widehat{R}^+$}{
    \If{$R \geq \alpha$
  }{
    \text{break}
  }
  }
    return $\hat{\lambda}, \alpha, 1-\delta$ 
  }{ \texttt{/*Risk-Confidence Correction*/}\\
  $\alpha_c = \min(\widehat{R}^+)$\\
  \For{$\delta_c\gets\delta$ \KwTo $0$ \KwBy $-10^{-2}$}{
  
  \For{$\hat{\lambda}, L$ \KwIn $\Lambda, L'$}{
    \If{$\textsc{WSR}(L, \delta_c) \leq \alpha$
  }{
    \text{break}
  }
  }
  }
  return $\hat{\lambda}, \alpha_c, 1-\delta_c$
  }
\end{algorithm*}

\end{document}